\documentclass[final,3p,times,2023]{elsarticle}

\usepackage{amssymb}
\usepackage{polyglossia}
\usepackage[dvipsnames]{xcolor}
\usepackage{array}
\usepackage{subfig}
\usepackage{amsmath}
\setdefaultlanguage{english}
\setotherlanguage{bengali}
\usepackage[hidelinks]{hyperref}
\newfontfamily\bengalifont[Script=Bengali]{Kalpurush.ttf}

\journal{Arxiv.org}

\begin{document}

\begin{frontmatter}



\title{IsharaKotha: A Comprehensive Avatar-based Bangla Sign Language Corpus}
\author[1]{MD. Ashikul Islam}
\ead{ashibul03abir@gmail.com}
            
\author[1]{Prato Dewan}
\ead{pratodewan1@gmail.com}
            
\author[1]{Md Fuadul Islam}
\ead{mdfuadulislam0@gmail.com}

\author[1]{Md. Ataullha}
\ead{ataullha00@gmail.com}
            
\author[1]{M. Shahidur Rahman \corref{cor}}
\ead{rahmanms@sust.edu}
\affiliation[1]{organization={Shahjalal University of Science \& Technology}, 
            city={Sylhet},
            postcode={3114},
            country={Bangladesh}}
\cortext[cor]{Corresponding author}


\begin{abstract}
Sign Language is an essential mode of communication for the hearing-impaired community, enabling them to interact and express themselves effectively. To overcome the gap that exists between the hearing and hearing-impaired communities, there is a need for a translation system that can convert text into sign language. Such a system can also be utilized to educate individuals interested in acquiring sign language skills. In order to create a sign language generation system, a sign language corpus is a must. In this article, the first-ever HamNoSys-based Bangla Sign Language corpus, \textit{IsharaKotha}, is presented for 3823 words. To enable the generation of signs from complete sentences, a deep learning-based lemmatizer was integrated to extract root words. An interface was subsequently developed to assess the quality of sign animations for letters, digits, and sentences. The evaluation involved two professional interpreters and a real sign language user, a hearing-impaired athlete, who rated the animations using categorical numeric scores. The average score achieved was 3.14 out of 4.00, indicating a high level of quality, categorized between ``Good'' to ``Excellent''. This outcome creates new opportunities for researchers to further develop and expand into dynamic translation systems utilizing this corpus. The evaluation system is accessible via the following link: \textit{\href{http://bdsl-isharakotha.ap-1.evennode.com}{http://bdsl-isharakotha.ap-1.evennode.com}}.
\end{abstract}




\end{frontmatter}

\section{Introduction}
\label{}
Sign Language (SL) is a language that is formed using hand, body movements and facial expressions. It is the only means of communication for hearing-impaired individuals. It allows them to interact with others and express themselves in ways that would be impossible through spoken language. Each nation possesses its own Sign Language (SL), each characterized by distinct syntax and structure.  For example, American Sign Language (ASL) is used primarily in the United States, while British Sign Language (BSL) is used in the United Kingdom. Similarly, the SL of Bangladesh is called Bangla Sign Language (BDSL).
\newline According to 2011 census data, around 13.7 million people, which accounts for approximately 9.6\% of the Bangladesh population, are either deaf or hard of hearing with a loss of 40 dB or more. Additionally, about 34.6\% of the population (49.2 million) has some kind of hearing loss, while 1.2\% of the population (1.7 million) suffers from severe hearing loss \cite{wikipediaDeafnessBangladesh}. The deaf community in Bangladesh is struggling due to the lack of proper SL education, which is resulting difficulties in their daily lives. Engaging interpreters are apparently the main solution to overcome the barriers that exist between normal and hearing impaired people. Unfortunately, such interpreters are scarcely available in our country.
\newline The Centre for Disability in Development (CDD), a non-profitable organization, took the initiative to document and publish manuals on BDSL on the basis of their experience of implementing development programs for persons with disabilities from 2000 \cite{cddCENTREDISABILITY}. Recently, following a series of workshops involving representatives from the Deaf and hearing-impaired community, a set of nearly 4000 signs was standardized and officially adopted by the Bangladesh Standards and Testing Institution (BSTI). This standardization was finalized under the guidance of the Bangladesh Computer Council (BCC), a statutory body within the ICT Division of the Ministry of Posts and Telecommunication \cite{bcc}.
Hence, those standardized signs have been used to create the proposed BDSL corpus.
\newline The majority of people in our society do not understand the meaning of sign gestures, just as deaf individuals may struggle with spoken languages. This leads to a persistent sense of isolation and numerous social challenges in their daily lives. Furthermore, members of this community have often faced discouragement from some individuals in society who hold the misguided belief that deafness or any form of disability is a form of punishment for past wrongs. To alleviate these communication barriers, an automated system is needed that can proficiently and reliably generate Bangla Sign Language (BDSL). 
\newline One of the easiest methods for learning SL is through the use of Sign Language generation systems, which employ video clips. But there are some disadvantages associated with this approach, such as the requirement for large memory in the user's system, expensive video clips, and time-consuming processes. As a result, animated SL generation systems have become more widely used and preferred due to their advantages over video-based systems. Although some works have been reported for interpreting Alphabets, Numerals and Words \cite{hoque2016automated,podder2022bangla,app}, these systems are inefficient as pre-recorded videos are used to generate the sign. In contrast, recent advances in Natural Language Processing (NLP) and computer graphics technologies have significantly improved the accuracy and effectiveness of avatar-based sign language translation systems. Avatars can translate text into sign language almost instantaneously. Therefore, these systems can be integrated into educational platforms, making online courses and educational materials more accessible to hearing-impaired students. Furthermore, such systems can provide visual and interactive support for learning the gestures and expressions associated with different words, phrases, and sentences making them well-suited for educating sign language to individuals with normal hearing. In this study,  we focused primarily on creating \textit{IsharaKotha}, a HamNoSys-based BDSL corpus for 3823 signs. An LSTM (Long Short-Term Memory) based lemmatizer has been introduced to convert complete sentences into root words, utilizing a self-developed corpus of 94,781 Bangla inflected words. Additionally, an interface has been developed to evaluate avatar-based animations for letters, digits, words, and complete sentences. Two professional interpreters and one sign language user conducted a total of 3,828 evaluations on randomly selected letters, digits, words, and complete sentences. Using a categorical numeric scoring system ranging from 1 (``Bad'') to 4 (``Excellent''), the evaluations resulted in an average score of 3.14 out of 4.00, indicating the high quality of the generated animations.

This research article is organized into 11 sections. Section 2 provides a review of related works in Bangla and other sign languages. Sections 3 and 4 describe the components of BDSL and the resources used in the creation of the BDSL corpus, respectively. Section 5 highlights the essential resources employed to generate avatar-based sign animations. Section 6 outlines the methodology adopted in this study, while Section 7 demonstrates how the corpus is utilized to translate Bangla text into animated sign language. The evaluation system used to assess the corpus is presented in Section 8. Section 9 presents the results obtained from the evaluation process, while Section 10 discusses the limitations of the study. Finally, Section 11 concludes the paper.

\section{Related Works}
\label{}
Numerous efforts have been made globally to develop sign dictionaries, which are specifically designed for Sign Languages used in various countries such as America, Spain, Britain etc. The creation of such dictionaries involves the utilization of either human videos or 3D avatars. A Swedish Sign Language Corpus (SSLC) was created by \cite{mesch2015gloss}, showcasing the video-recorded discussions with 42 participants. They discussed about gloss annotation scheme used for lexical signs, finger spelling, and productive signs, with ongoing efforts to annotate the entire corpus, covering both one- and two-handed signs based on a collection of 33,600 tokens from the SSLC. Gutierrez-Sigut \cite{gutierrez2016lse} built a lexical database for Spanish Sign Language, which comprises a free online tool consisting of 2400 videos. 
A Japanese Sign Language dictionary was generated for medical purposes by Nagashima et al. \cite{nagashima2016support}, featuring 3D animations to demonstrate sign motions where two types of search options are available: searching by the description of signs and searching by Japanese keywords. In a study conducted by Goyal \cite{goyal2016development}, a synthetic animation dictionary was prepared for Indian Sign Language(ISL), which generated a total of 1818 signs. The generated signs were created using the Sign Editor software and then converted into SiGML files. 
A comprehensive collection of nearly 15 hours of annotated video material from 21 Finnish Sign Language (FinSL) users was organized by Salonen et al. \cite{salonen2020corpus}. It includes the annotations of ID-glosses, Finnish translations, and metadata organized according to the IMDI standard. Sugandhi et al. \cite{sugandhi2020sign} proposed an avatar-based dictionary, creating a multilingual ISL with Hamburg notation. Additionally, a 3D animated dictionary of Arabic Sign Language (ArSL) containing about 3000 signs was constructed by Aliwy et al. \cite{aliwy2021development}, which is the largest HamNoSys-based corpus for ArSL.\\
While many languages have made significant progress in developing sign generation systems, efforts to computerize BDSL are still in their early stages. The first isolated character dataset of images for BDSL was produced by Islam et al. \cite{islam2018ishara} which consists of 50 sets of 36 Bangla basic sign characters, collected with the assistance of deaf and non-deaf volunteers. Several studies have focused on character recognition \cite{shamimul2023multi,al2023real,akash2023action,khatun2021systematic}. Islam et al. \cite{islam2022automatic} trained a model only for 13 Bangla Numeral gestures using HamNoSys. A video based sign language dataset is also presented by Sams et al. \cite{sams2023signbd}. More recently, Hasib et al. \cite{hasib2022bdsl} developed a dataset \textit{BDSL}49, which includes a total of 29,490 images depicting hand signs for the 49 Bangla alphabets. Subsequently, machine learning models have been employed to identify these Bangla alphabets.
To our knowledge,  no work has been undertaken in translating text to BDSL that is suitable for real world communication. Since a corpus is a prerequisite for developing such a translation system, a comprehensive avatar-based  BDSL corpus \textit{IsharaKotha} has been created by using HamNoSys notations. HamNoSys notations offer flexibility in capturing various facets of sign language and this advantage has been chosen for producing the proposed avatar-based animations for BDSL corpus. 

\section{Bangla Sign Language (BDSL)}
Effective and accurate use of SL requires a proper understanding of its linguistic features and principles. A user guideline ``Specification of Bangla Sign Language (first version)" \cite{corpusDoc} is collected personally from the  Standardization  committee of Bangla Language for use in Information Technology. According to the user manual, the components of BDSL can be divided into two categories:
\begin{itemize}
    \item[-] Characteristics 
    \item[-] Primary Elements
\end{itemize}
\subsection{Characteristics}
The characteristics component consists of 3 parts: Gestures, Lip Reading and Facial Expression.
\begin{itemize}
\item \textbf{Gestures:} 
Sign Languages use different types of gestures to convey meaning. For instance, nodding the head up and down indicates ``yes" and side to side indicates ``no".
\item \textbf{Lip Reading:}
Lip reading is the ability to understand spoken language by observing the movements of a speaker's lips without hearing the words.
\item \textbf{Facial expression:}
Facial expressions play an important role in communication, whether it is spoken language or SL. These expressions convey a person's emotions, such as happiness, sadness, surprise, fear and expressiveness. Expressions like widening the eyes, frowning, or puckering lips can effectively express emotions and enhance communication. Therefore, proper use of facial expressions is vital for a meaningful conversation.
\end{itemize}

\subsection{Primary Elements}
There are five fundamental elements of BDSL. These are:
\begin{enumerate}
\item \textbf{Hand Shape:}
There are more than 30 different hand shapes required for representing BDSL. Each representation requires a different hand shape. Depending on the form, different names are given to hand shapes.
\item \textbf{Direction of Hand:}
Direction is another important element of SL. Some hand shapes need multiple fingers and their tip positions may vary. Depending on which finger is used, the location of that finger's tip will determine the direction of the hand.
\item \textbf{Position of Palm:}
Another major element of SL is the palm. Variations in hand shape and direction result in different signs. Similar to this, different palm position also represents different signs.
\item \textbf{Movement of hand:}
Hand movements are very important in SL. Hands may remain still in representing many gestures. Again, in many cases, the hands need to move in different directions one or more times.
\item \textbf{Position of Hand:}
Finally, proper hand position is essential for accurate expression in SL. Hand position refers to the height at which the hand will be positioned, such as the waist, the shoulders, the chest and the forehead.
\end{enumerate}
\section{An Overview of BDSL Dictionary}
The recently adopted BDSL dictionary encompasses nearly four thousand signs, representing words from a diverse range of categories: \textbengali{পরিবার ও আত্মীয়} (Family \& Relatives), \textbengali{সম্পর্ক} (Relationships), \textbengali{শিক্ষা} (Education), \textbengali{কাজ} (Job), \textbengali{মানুষের চারিত্রিক বৈশিষ্ট্য} 
\begin{figure}[h]
    \centering
    \subfloat{
        \includegraphics[scale=.6]{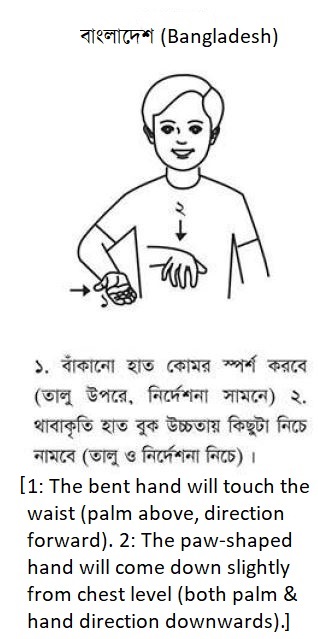}}
    \subfloat{
        \includegraphics[scale=.6]{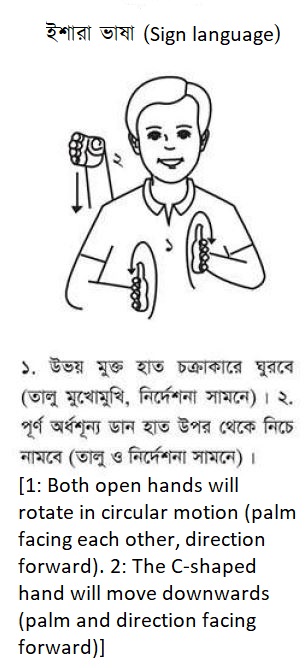}}\hspace{.05cm}
    \subfloat{
        \includegraphics[scale=.6]{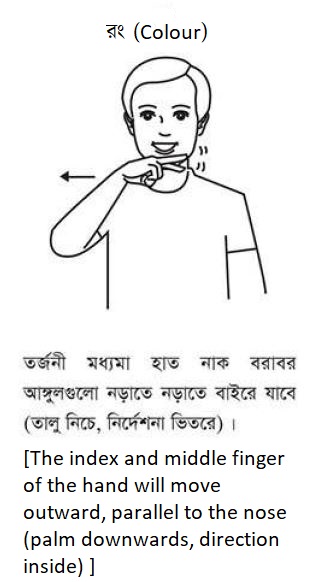}}
    \subfloat{
        \includegraphics[scale=.6]{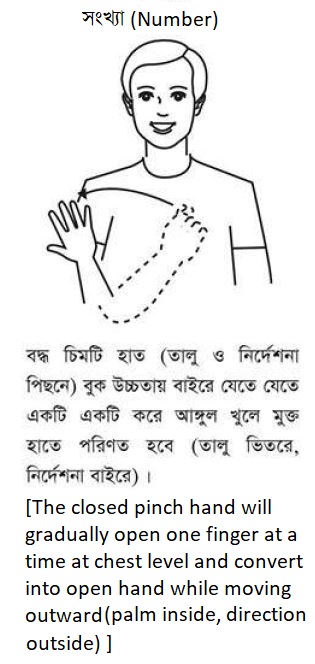}}\hspace{.05cm}
    \caption{Example Signs from BDSL Dictionary}
    \label{Dataset}
\end{figure}
(Human Characteristics), \textbengali{প্রকৃতি ও পরিবেশ} (Nature \& Environment), \textbengali{ভৌগলিক উপাদান} (Geographical Elements), \textbengali{রোগ ও চিকিৎসা} (Disease \& Treatment), \textbengali{খেলাধুলা} (Sports), \textbengali{ধর্ম} (Religion), \textbengali{প্রাতিষ্ঠানিক শব্দাবলী} (Institutional Words), \textbengali{সামাজিক রীতিনীতি ও অনুষ্ঠান} (Social Custom \& Occasion) etc.
The words above the figure indicate the name of the sign language word, while the sentences below describe how to create the respective sign.
\section{Creating Avatar-based Animations for Sign Language}
Avatar-based animations provide an efficient and effective means to translate texts into sign language. The increasing popularity of avatar-based translation systems reflect a positive step towards a more inclusive and accessible world for individuals who rely on sign language for communication. This section will provide a brief overview of the process involved in creating animations using avatars. 
\subsection{HamNoSys}
In 1984, Siegmund Prillwitz developed the HamNoSys notation. It is a standard transcription system for expressing SL that has nearly 200 characters and can be used anywhere \cite{hanke2004hamnosys}. It has both non-manual and manual parts of the body. An example of a phonetic notation is HamNoSys, which comprises various factors that are referred to as handshape, orientation, location and movement.
\begin{itemize}
    \item \textbf{Handshape:}
    There are many different handshape symbols available in HamNoSys for representing signs as shown in Figure~\ref{fig: Handshape symbols}.
\begin{figure}[h!]
      \centering
      \includegraphics[width=0.55\textwidth]{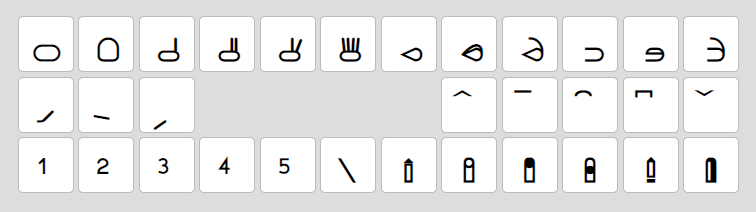}
      \caption{Handshape symbols}
      \label{fig: Handshape symbols}
\end{figure}

    \item \textbf{Orientation:}
    The hand orientation symbols define the direction of fingers and palm as shown in Figure~\ref{fig: Orientation symbols}. There are symbols available for six different directions.
\begin{figure}[h!]
      \centering
      \includegraphics[width=0.45\textwidth]{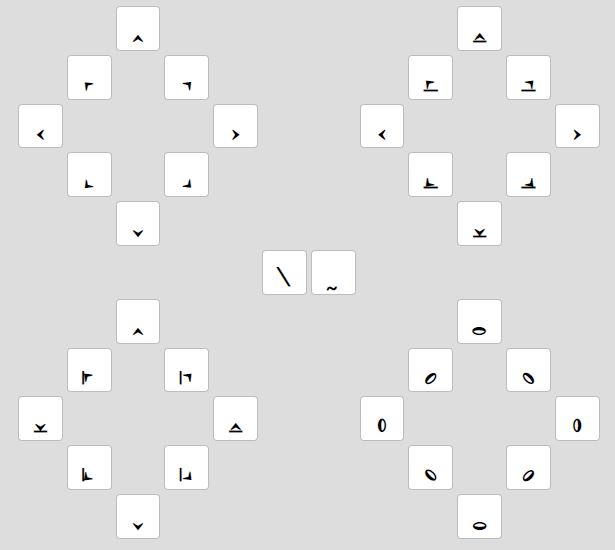}
      \caption{Hand Orientation symbols}
      \label{fig: Orientation symbols}
\end{figure}

    \item \textbf{Location:}
    The location symbols show where the hand will be placed on the body. There are more than 30 symbols for representing hand location. The symbols used for representing locations are shown in Figure~\ref{fig: Location symbols}.
\begin{figure}[h!]
      \centering
      \includegraphics[width=0.53\textwidth]{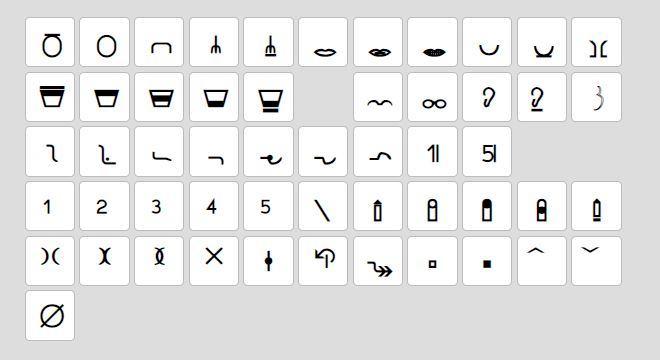}
      \caption{Location symbols}
      \label{fig: Location symbols}
\end{figure}

    \item \textbf{Movement:}
    In HamNoSys, there is a range of rotational and directional symbols available. These symbols can be observed in Figure~ \ref{Fig:Straight Movement Section} for straight movement and Figure~\ref{Fig:Circular Movement Section} for circular movement, respectively. The movements can also be repeated.
\begin{figure}[!htb]
   \begin{minipage}{0.5\textwidth}
     \centering
     \includegraphics[width=.8\linewidth]{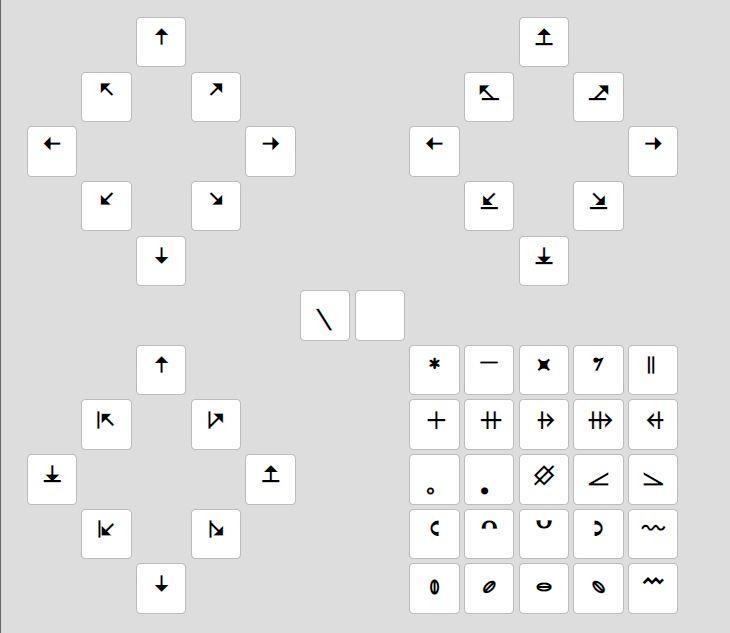}
     \caption{Straight movement symbols}\label{Fig:Straight Movement Section}
   \end{minipage}\hfill
   \begin{minipage}{0.5\textwidth}
     \centering
     \includegraphics[width=.7\linewidth]{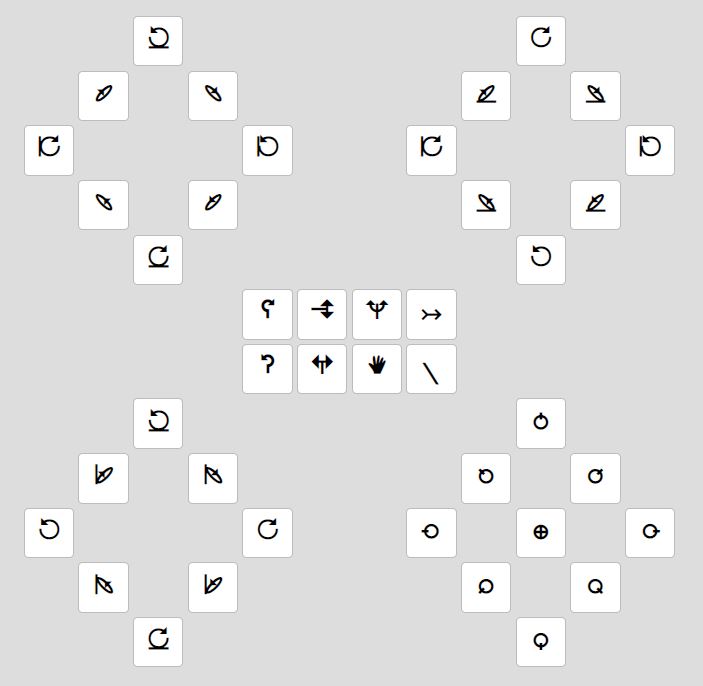}
    \caption{Circular movement symbols}\label{Fig:Circular Movement Section}   
    \end{minipage}
\end{figure}

    \item \textbf{Others:}
    There are extra symbols available in HamNoSys for two-handed movements. It also provides symbols for non-manual features like head and facial gestures, movement of lips and eyes.
\begin{figure}[h!]
      \centering
      \includegraphics[width=0.5\textwidth]{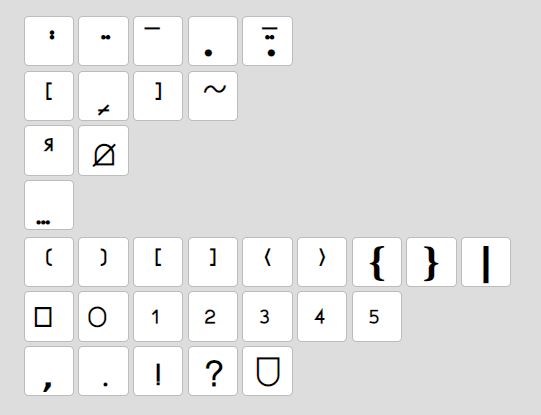}
      \caption{Two-Hand based symbols}
      \label{fig: Two-Handed Section}
\end{figure} 

\end{itemize}
\subsection{SiGML}
SiGML is a type of XML scripting language that is used to represent each symbol in HamNoSys for a particular sign \cite{elliott2004overview}. It was first developed in ViSiCAST project \cite{bangham2000virtual} to provide signing gesture interpretation in a way that makes it possible to animate those gestures in real-time using an avatar. However, a ``gloss" attribute has a meaning that frequently relates to the name of the word for which its SiGML is generated. The non-manual features can also be included. The equivalent tags for the non-manual features are placed between the tags \textit{<hamnosys\_nonmanual>} and \textit{</hamnosys\_nonmanual>}. On the other hand, the tags \textit{<hamnosys\_manual>} and \textit{</hamnosys\_manual>} are used for manual ones.  
\subsection{JASigning Avatars}
JASigning (Java Avatar Signing) is a software system that generates synthetic SL animations using a virtual human or avatar \cite{elliott:10035:sign-lang:lrec}. The system is designed to work with both desktop and web-based applications, allowing users to input SL sequences using the SiGML notation. The system is capable of operating in real-time, meaning that the avatar can generate and perform SL sequences in response to user interaction. This system provides users with a choice of five distinct standard avatars to visually represent SL. 

\section{Development of the \textit{IsharaKotha} Corpus}

\begin{figure}[h!]
      \centering
      \includegraphics[width=0.65\textwidth]{"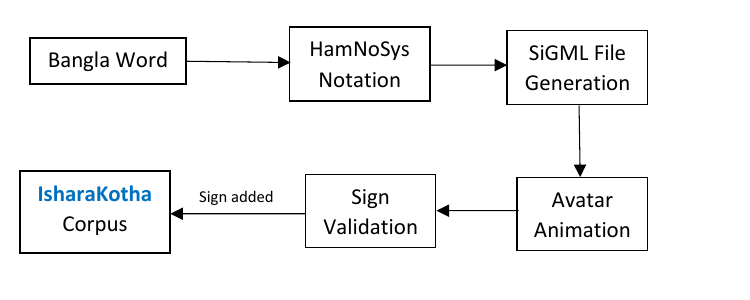"}
      \caption{Workflow of generating the \textit{IsharaKotha} Corpus}
      \label{fig: Corpus}
\end{figure}

The process for developing the \textit{IsharaKotha} corpus has been illustrated in Figure~\ref{fig: Corpus}.

\subsection{HamNoSys Notation Creation}
\begin{figure}[h!]
      \centering
      \includegraphics[width=0.87\textwidth]{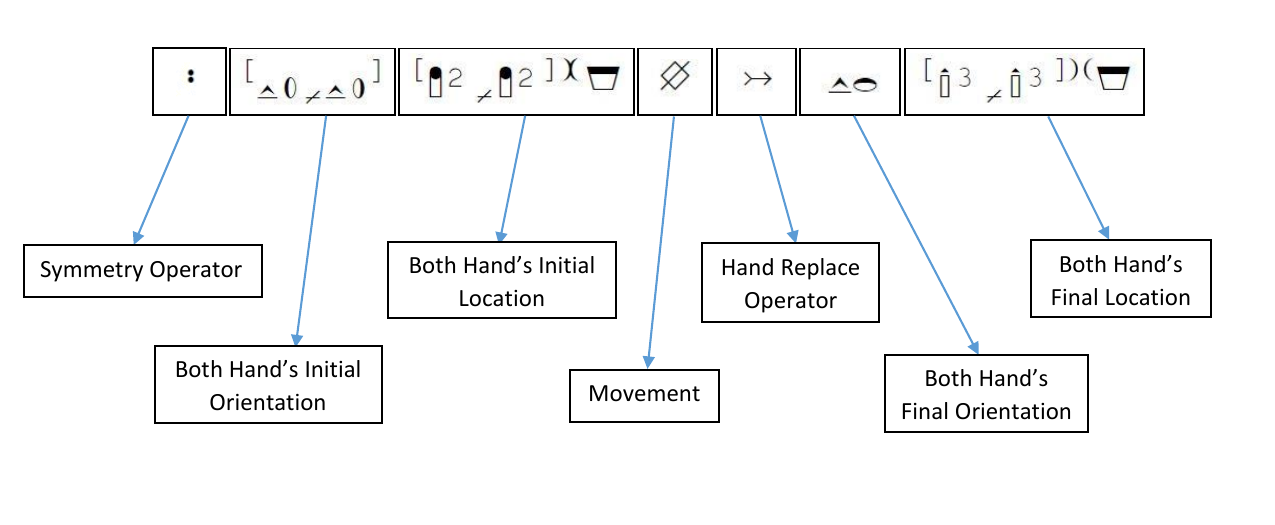}
      \caption{HamNoSys Notation for \textbengali{বই} (Book).}
      \label{fig:Boi HamNoSys}
\end{figure}

To create the animation of a word, the initial step involves generating the suitable HamNoSys notation for that specific word. To accomplish this, an open-source available software called ``HamNoSys" \cite{dgs_korpus_ap04_2021_02} has been utilized. This software offers flexibility in configuring various parameters, encompassing aspects such as hand gestures, movements, positions, and even head and body gestures. These parameters encompass handshape, hand orientation, position, and movement among others.
Figure~\ref{fig:Boi HamNoSys} describes the symbols used to represent the sign for \textbengali{বই} (Book).
In order to express the Bangla word \textbengali{`বই'}(Book), it is necessary to utilize a symmetry operator that signifies the involvement of both hands. Initially, both hands are oriented in the same position but in opposite directions. Moreover, the index fingers of both hands are touching each other as they are clasped together across the chest. Subsequently, a slight modification to the hand configuration is made to ensure the precise representation of the sign, where both hands are located to each other at a previously defined position.
    
\begin{figure}[h]
    \centering
    
    \subfloat[SiGML file of the word \textbengali{বই} (Book).]{
        \includegraphics[width=0.48\textwidth,height=9.3cm]{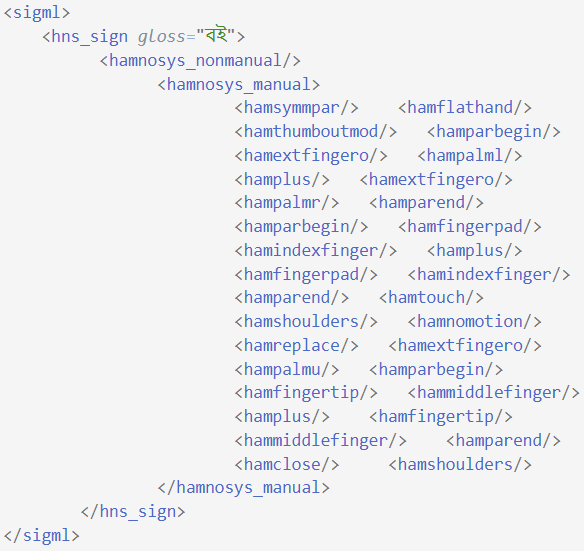}
        \label{fig:sigml_boi}
    }
    \hfill
    \subfloat[Avatar animation of \textbengali{বই} (Book).]{
        \includegraphics[scale=.3]{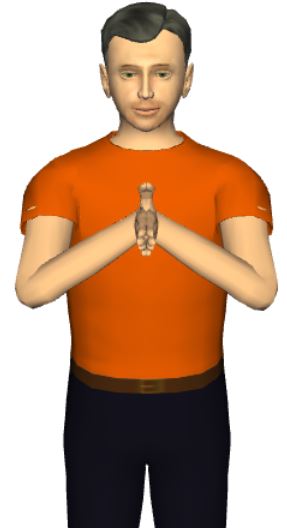}
        \includegraphics[scale=.3]{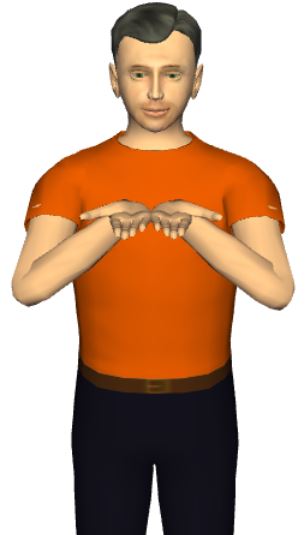}
        \label{fig:animation_boi}
    }

    \caption{Visualization of SiGML description and avatar animation for the word \textbengali{বই}.}
    \label{fig:combined_boi}
\end{figure}

\subsection{HamNoSys to SiGML file generation}
After generating the appropriate HamNoSys notation for a particular word, the SiGML files are needed to construct to animate the sign of the word. A publicly accessible web platform, ``SiS-Builder" have been used to create the relevant SiGML files \cite{goulas2010sis}.
The corresponding SiGML file of the word \textbengali{বই} (Book) is shown in Figure~\ref{fig: Sigml for boi}.

\subsection{SiGML to Animation Creation}
To see how the SiGML file for a specific file works, a software called SiGML player \cite{ueaCWASAVhg2023}, available for Windows, Mac and Linux operating systems has been utilized. It provides 3D avatar characters to show the visual representation of a sign. Moreover, this tool allows us to input the SiGML and view the corresponding animation. The avatar animation of \textbengali{'বই'} (Book) is provided in Figure~\ref{Animation for boi}. Once the correct animation is produced, the SiGML file is included in the corpus.

The proposed \textit{IsharaKotha} corpus consists SiGML files of 3823 sign language words. The details of our corpus across different categories are outlined in Table~\ref{Corpus Description}.
\begin{table}[h!]
\centering
\caption{Description of the \textit{IsharaKotha} Corpus}
\renewcommand{\arraystretch}{1.15}
\begin{tabular}{|c|m{20em}|c|} 
\hline
Index &\centering Category & Word Count\\
\hline
1 & \centering \textbengali{অপরাধ ও আইন} (Crime \& Law) & 38 \\
2 & \centering \textbengali{অর্থনৈতিক} (Economics) & 35 \\
3 & \centering \textbengali{আন্তর্জাতিক} (International) & 26 \\
4 & \centering \textbengali{উৎসব} (Festival) & 21 \\
5 & \centering \textbengali{ঐতিহাসিক} (Historical) & 11 \\
6 & \centering \textbengali{কর্মী} (Worker) & 77 \\
7 & \centering \textbengali{কাজ} (Job) & 45 \\
8 & \centering \textbengali{খাবার ও পানীয়} (Food \& Drinks) & 234 \\
9 & \centering \textbengali{গুরুত্বপূর্ণ স্থান} (Important Places) & 20 \\
10 & \centering \textbengali{গৃহস্থালী সামগ্রী} (Household Items) & 342 \\ 
11 & \centering \textbengali{দেশ} (Country) & 39 \\
12 & \centering \textbengali{ধর্ম} (Religion) & 105 \\
13 & \centering \textbengali{পরিবহন} (Transportation) & 91 \\
14 & \centering \textbengali{পরিবার ও আত্মীয়} (Family \& Relatives) & 69 \\
15 & \centering \textbengali{পশুপাখি} (Animals \& Birds) & 121 \\
16 & \centering \textbengali{পেশা} (Profession) & 65 \\
17 & \centering \textbengali{পোষাক পরিচ্ছেদ} (Clothing) & 120 \\
18 & \centering \textbengali{প্রকৃতি ও পরিবেশ} (Nature \& Environment) & 133 \\
19&\centering \textbengali{অক্ষর} (Alphabets) & 49 \\
20&\centering \textbengali{ভূগোল} (Geography)& 19 \\
21&\centering \textbengali{ব্যাকরণ} (Grammar)& 21 \\
22&\centering \textbengali{মানুষের চারিত্রিক বৈশিষ্ট্য} (Human Characteristics) & 470 \\
23&\centering \textbengali{মানুষের শরীরের অঙ্গ} (Human Body Parts)& 67 \\
24&\centering \textbengali{রাজনীতি} (Politics)& 149 \\
25&\centering \textbengali{রোগ ও চিকিৎসা} (Disease \& Treatment) & 84 \\
26&\centering \textbengali{শহর} (City)& 63 \\
27&\centering \textbengali{শিক্ষা} (Education)& 269 \\
28&\centering \textbengali{সংখ্যা} (Number)& 10 \\
29&\centering \textbengali{সামাজিক কাজ} (Social Work)& 54 \\
30&\centering \textbengali{বিনোদন} (Entertainment)& 28 \\
31&\centering \textbengali{যন্ত্র} (Machine)& 48 \\
32&\centering \textbengali{খেলা} (Sports)& 53 \\
33&\centering \textbengali{কৃষি} (Agriculture)& 30 \\
34&\centering \textbengali{রঙ} (Colour)& 14 \\
35&\centering \textbengali{সম্পর্ক} (Relation)& 27 \\
36&\centering \textbengali{অন্যান্য} (Others)& 776 \\ 
\hline
 &\centering \textbf{Total} & 3823 \\
\hline
\end{tabular}
\label{Corpus Description}
\end{table}
Some examples of the generated signs are provided in Table~\ref{Examples}.
\begin{table}[h!]
\centering
\caption{Examples of Generated Animations}
\fontsize{7}{8}\selectfont
\begin{tabular}{p{1.4cm} p{1.8cm} p{2.5cm} c p{3cm}}
\hline
Word &\centering BDSL Word &\centering Sign Description &\centering HamNoSys Notation & \hspace{.4cm} Avatar Animation \\
\hline 
\textbengali{মা}(Mother) & \raisebox{-0.15\height}{\includegraphics[scale = 0.035]{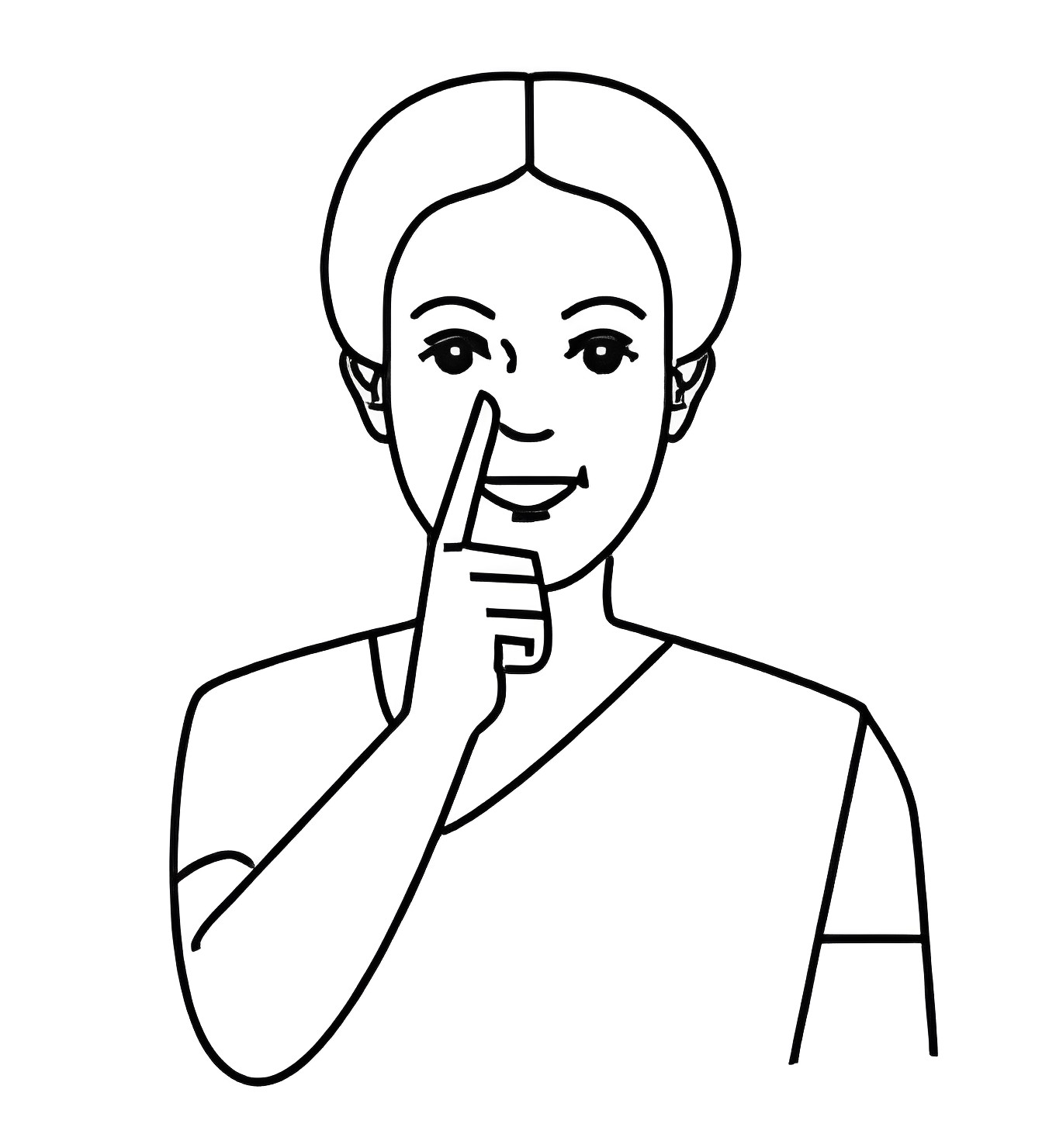}} & \vspace{-1.55cm} \textbengali{তর্জনী হাত নাক স্পর্শ করবে (তালু ভিতরে, নির্দেশনা উপরে)} [The index finger of right hand touches the nose (palm inside, direction up)] &  \includegraphics[scale=.5]{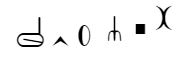}  &  \hspace{.8cm} \raisebox{-0.205\height} {\includegraphics[width=1.35cm,height=2cm]{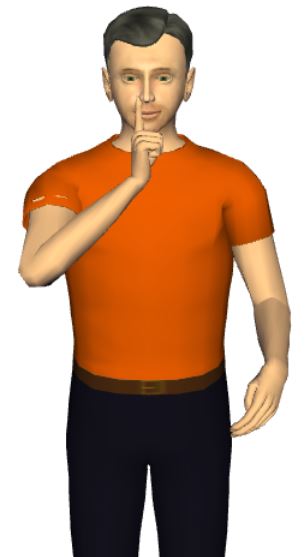}} \\ \\
 \textbengali{বাসা}(House) & \includegraphics[scale = 0.04]{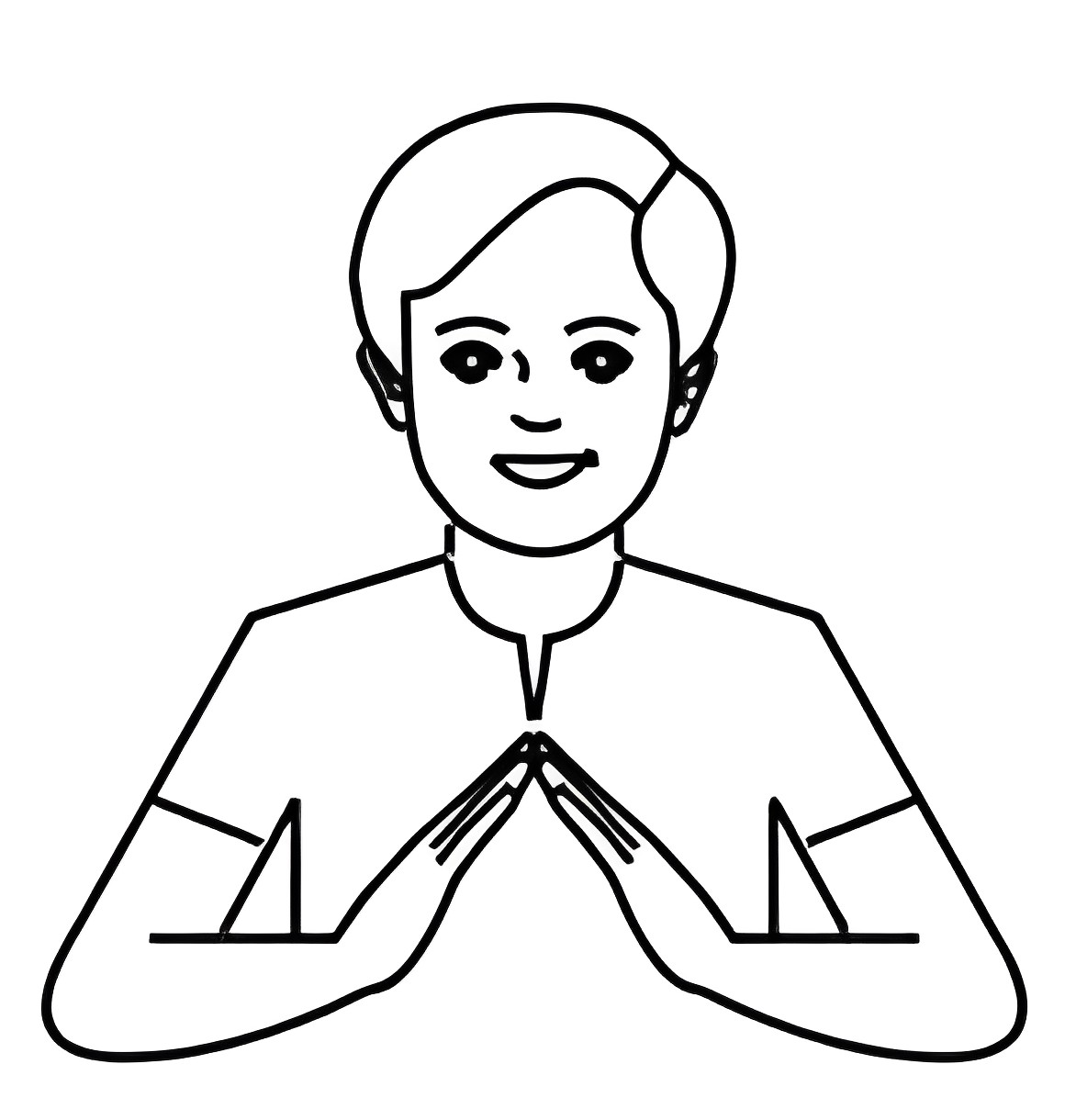} & \vspace{-1.55cm} \textbengali{উভয় চ্যাপ্টা হাত কৌণিকভাবে প্রান্তভাগ স্পর্শ করবে (তালু মুখোমুখি, নির্দেশনা উপরে)} [Both flat hands will touch the edge at an angle (palms facing each other, direction above)] &  \includegraphics[scale=.5]{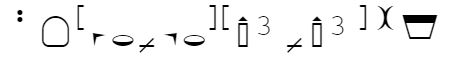}   & \hspace{.8cm} \raisebox{-0.25\height}{\includegraphics[width=1.35cm,height=2.25cm]{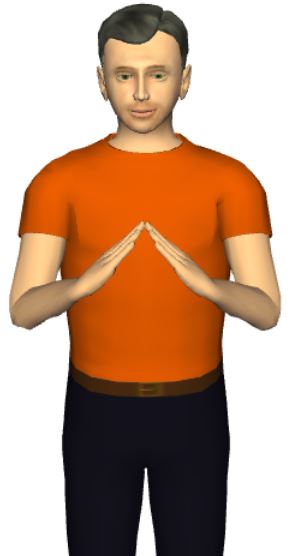}} \\ \\
\vspace{-.8cm} \textbengali{সাত}(Seven) & \includegraphics[scale = 0.04]{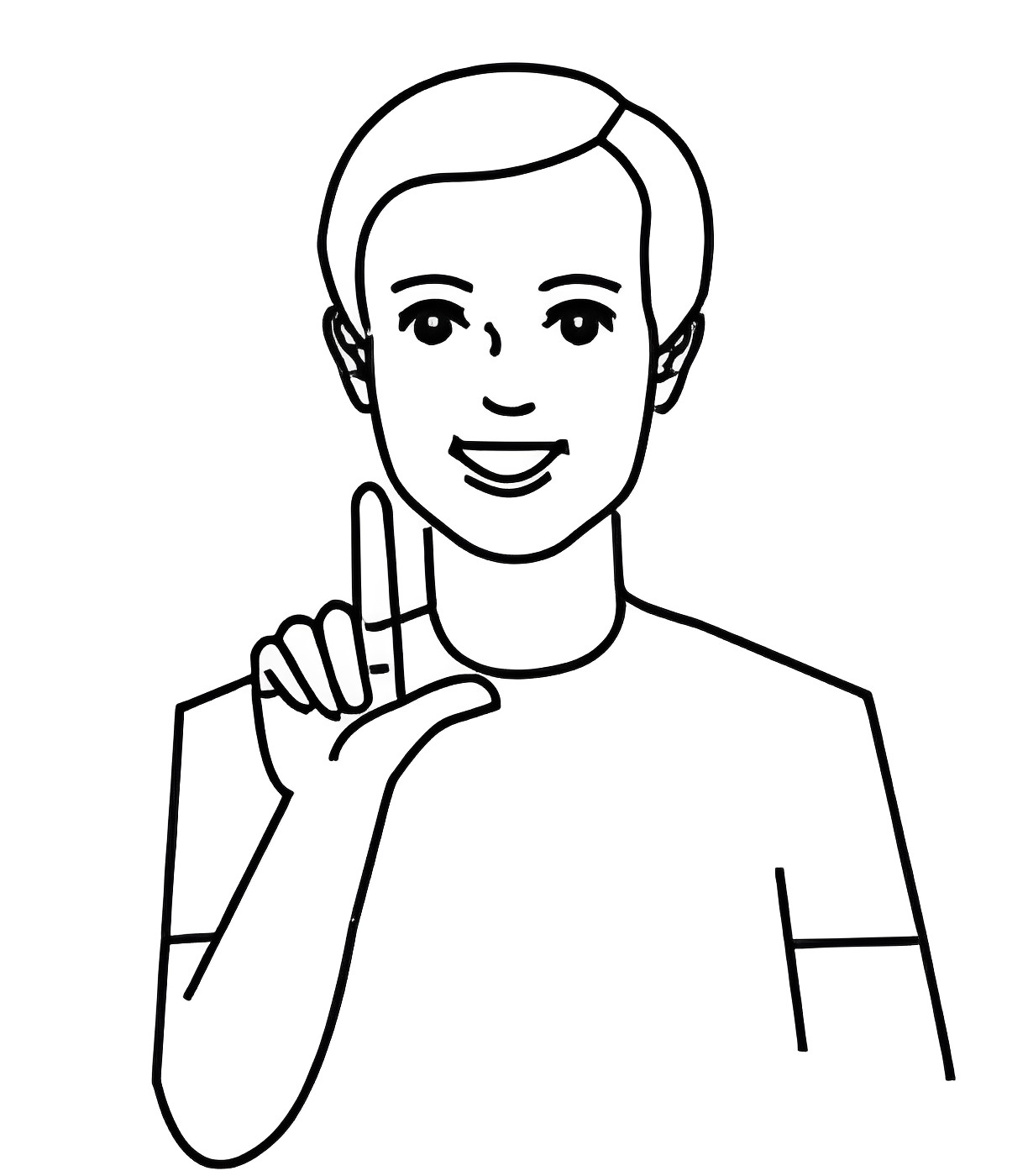} &  \vspace{-1.75cm}\textbengali{তর্জনী বৃদ্ধাঙ্গুলি হাত বুক উচ্চতায় স্থির থাকবে (তালু সামনে, নির্দেশনা উপরে)} [The forefinger and thumb are positioned at chest height (palm forward, direction above)] &  \includegraphics[scale=.5]{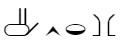}   & \hspace{.8cm} \raisebox{-.2\height}{\includegraphics[width=1.4cm,height=2.25cm]{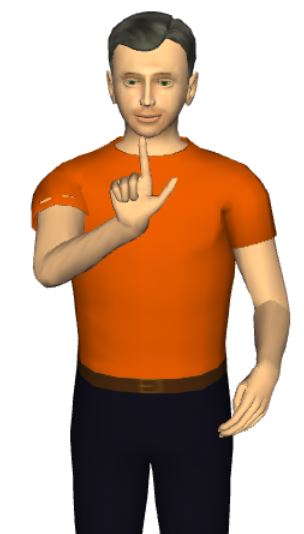}} \\ \\
\vspace{-.8cm} \textbengali{ধন্যবাদ}\newline (Thank You) & \includegraphics[scale = 0.039]{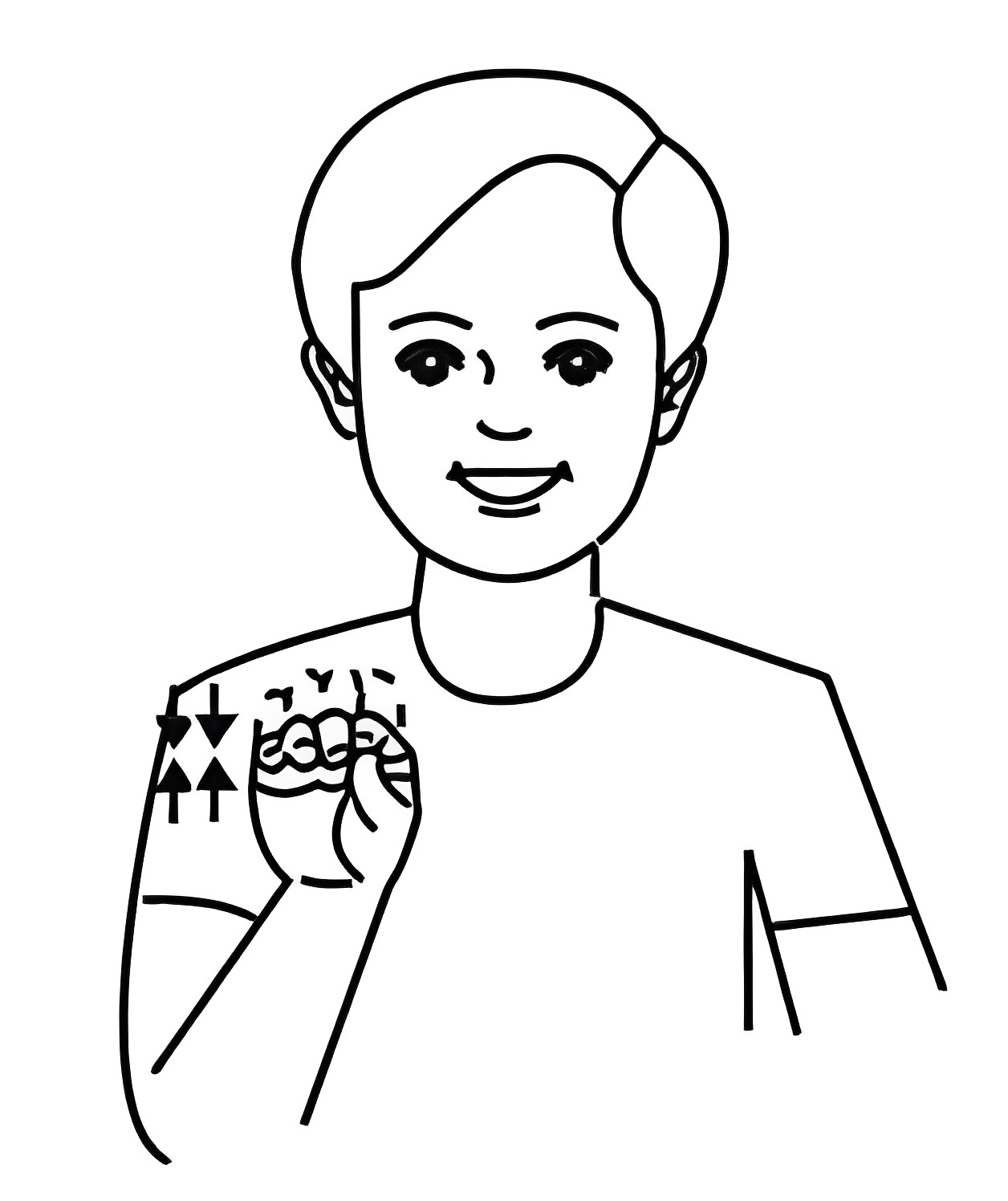} & \vspace{-2.1cm} \textbengali{সমান্তরাল ভাজ করা হাত ব্যক্তিকে নির্দেশ করে দু'বার বদ্ধ গোছাকৃতি হাতে পরিণত হবে (উভয়ক্ষেত্রে- তালু ও নির্দেশনা সামনে)} [Parallel folded hand will turn into fingers pinched handshape twice, indicating at the person(In both cases, palm and direction ahead)] &  \includegraphics[scale=.39]{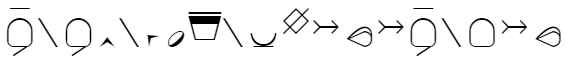}   &  \raisebox{-0.2\height}{\includegraphics[width=1.4cm,height=2.55cm]{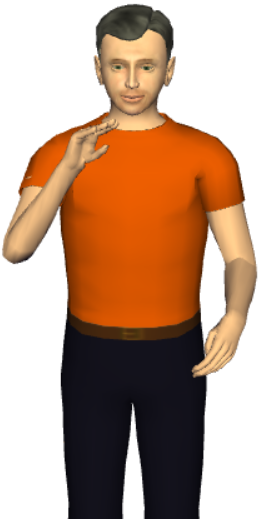}  \includegraphics[width=1.4cm,height=2.55cm]{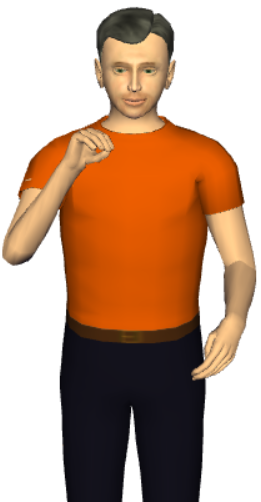} } \\ \\
\vspace{-.35cm} \textbengali{নাম}(Name) & \includegraphics[scale = 0.04]{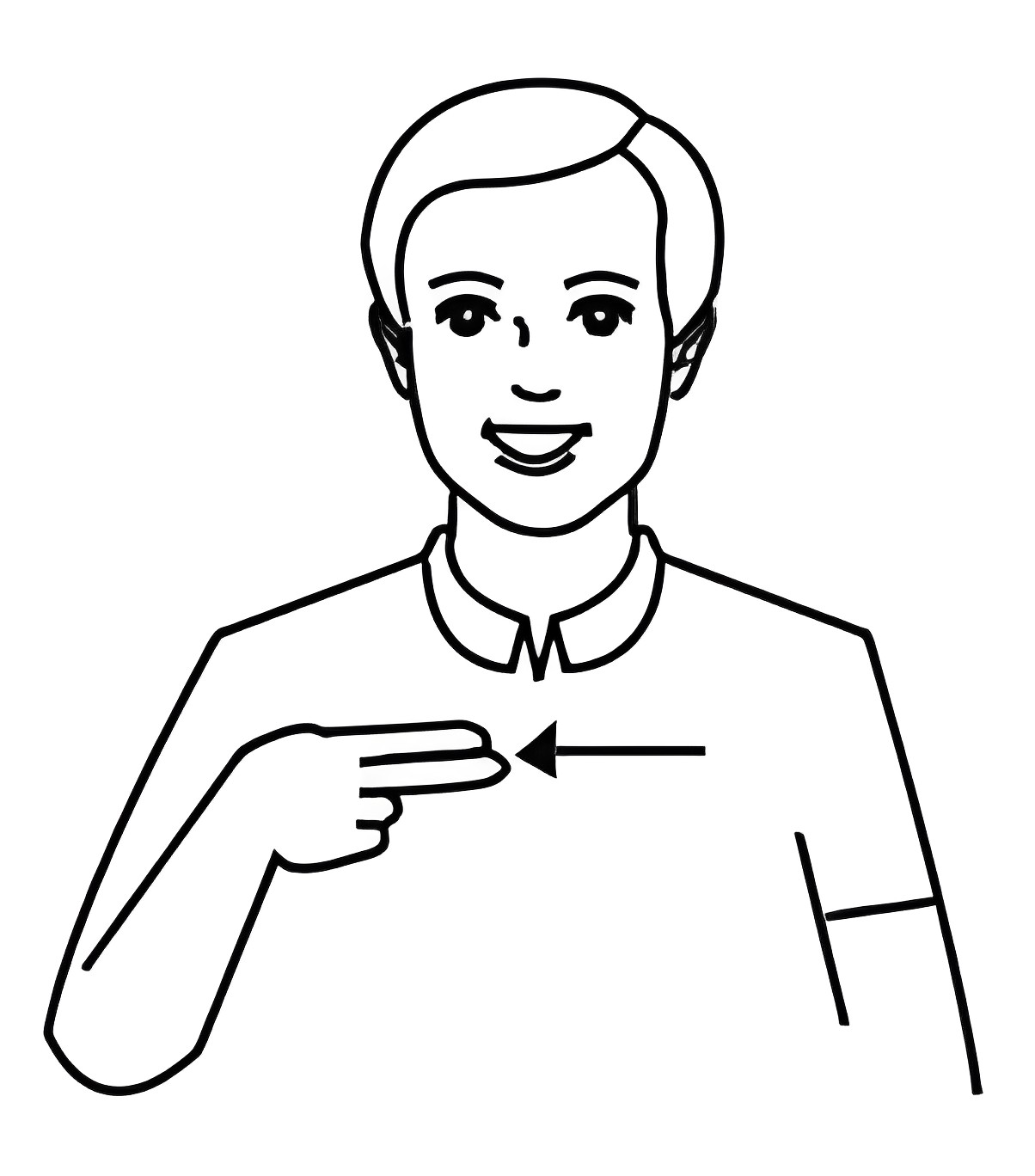} & \vspace{-2.2cm} \textbengali{তর্জনী মধ্যমা হাত বুক স্পর্শ করা অবস্থায় ভিতর থেকে বাইরে যাবে (তালু পিছনে, নির্দেশনা ভিতরে)} [The index and middle finger of the hand will move from inside to outside touching the chest (palm backward, direction inside)] & \includegraphics[scale=.4]{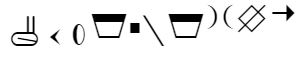}  & \includegraphics[width=1.4cm,height=2.55cm]{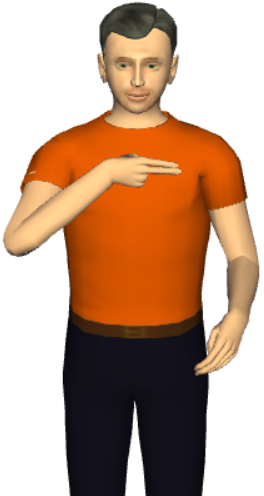} \hspace{-.10cm} \includegraphics[width=1.4cm,height=2.55cm]{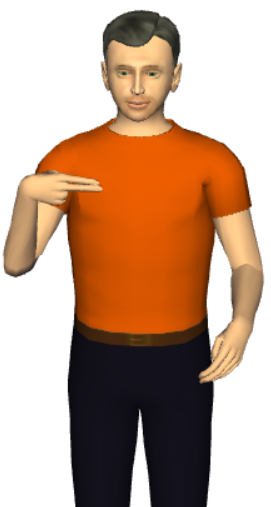} \\ \\
\hline
\end{tabular}
\label{Examples}
\end{table}
\clearpage

\section{Translating Bangla text to Avatar animation} 

The primary objective behind the creation of the \textit{IsharaKotha} Corpus is to facilitate the development of a Bangla text-to-sign animation generation system using avatars \cite{elliott:10035:sign-lang:lrec}.
As demonstrated in Figure~\ref{fig: system}, our text to animated sign language generation system workflow involves several key steps:
\begin{figure}[h!]
      \centering
      \includegraphics[width=0.7\textwidth]{"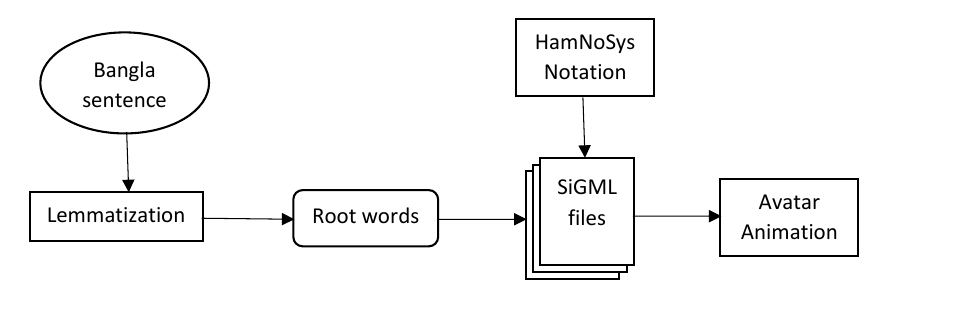"}
      \caption{Workflow of text-to-animated sign language generation}
      \label{fig: system}
\end{figure}

\begin{figure}[h!]
      \centering
      \includegraphics[width=0.5\textwidth]{"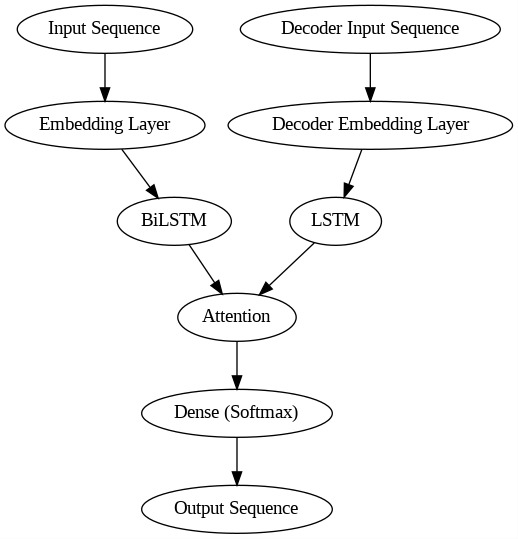"}
      \caption{Architecture of the Lemmatizer}
      \label{fig: Lemmatizer}
\end{figure}
\begin{itemize}
    \item[i)] \textbf{Input Processing:} Bangla sentences serve as our input data. 
    \item[ii)] \textbf{Lemmatization:} Lemmatization is the process of finding the root form of an inflected word. Our lemmatizer utilizes a sequence-to-sequence (Seq2Seq) architecture with an attention mechanism to transform inflected words into their lemmas. As shown in Figure~\ref{fig: Lemmatizer}, the model features a BiLSTM encoder that processes the input sequence bidirectionally, capturing both forward and backward context, and generates hidden states along with concatenated context vectors. These context vectors are then used to initialize the states of a unidirectional LSTM decoder, which produces the output sequence token by token.
An attention mechanism is incorporated between the encoder and decoder, enabling the decoder to focus on relevant parts of the input sequence at each step of the decoding process. The decoder's outputs, augmented with the attention context, are passed through a dense layer with softmax activation to predict the next token in the sequence. This design ensures efficient encoding of context and dynamic focus during decoding, resulting in accurate sequence generation.
The lemmatizer was trained and tested on a self-developed corpus containing 94,781 unique inflected Bangla words, achieving an accuracy of 79.22\%.
    
    \item[iii)] \textbf{SiGML Files Retrieval:} Next, we retrieve SiGML files corresponding to the extracted root words. Those files contain the necessary information for generating avatar animations. 
    \item[iv)] \textbf{Avatar Animation:} Finally, the retrieved SiGML files are utilized to animate avatars, thereby generating animations for complete Bangla sentences. It is noted that the currently our system can generate sign animations when all root words are available in the sign dictionary. The implementation of the dynamic system requires mapping the entire vocabulary to the developed corpus, a task that has not yet been accomplished.
\end{itemize}


\section{Evaluation of \textit{IsharaKotha} Corpus}
\begin{figure}[h!]
      \centering
      \includegraphics[width=0.35\textwidth]{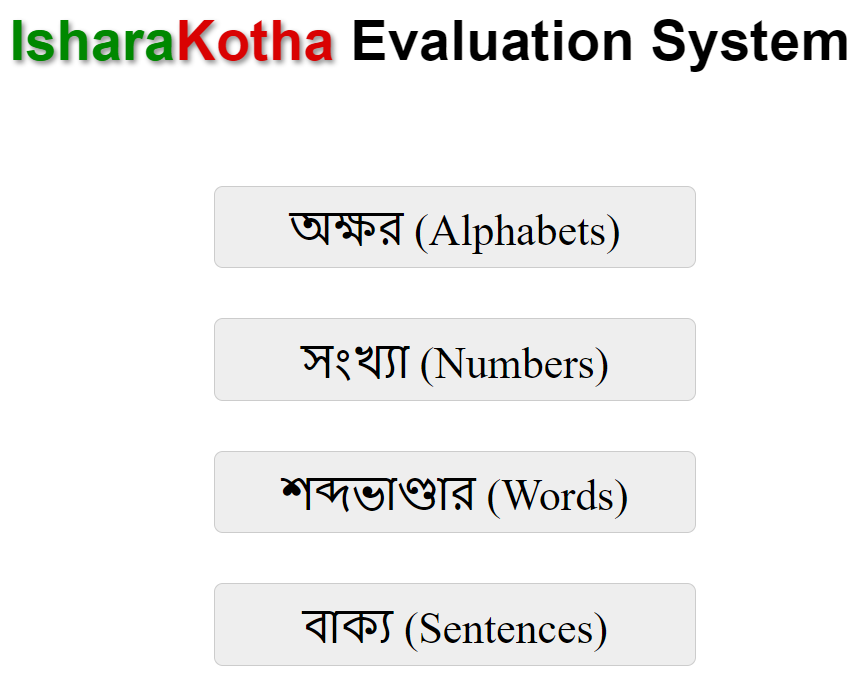}
      \caption{Sections in the Evaluation System}
      \label{fig: sections}
\end{figure}

A web interface has been created for evaluating the \textit{IsharaKotha} corpus. The main objective of this evaluation system is to gather valuable feedback from interpreters to assess the quality and effectiveness of the  corpus animations. The interface is divided into four main sections: Alphabets, Numbers, Words and Sentences, as depicted in Figure~\ref{fig: sections}.\\
In the ``Alphabets" section, individual alphabets in our corpus are displayed through animations using an avatar. Similarly, the ``Numbers" section showcases the representations of numerical symbols for interpreters to evaluate. The ``Words" section is further categorized into 28 categories like \textbengali{কাজ} (Work), \textbengali{শিক্ষা} (Education), \textbengali{অর্থনৈতিক} (Economy), \textbengali{দেশ} (Country), and more. Figures~14 and 15 provide a description of the word categories and the words within the \textbengali{কাজ} (Work) category, respectively. Interpreters can select a category of interest to explore animations for various words within that category.
Finally, the web interface encompasses a ``Sentences" section, as depicted in Figure~\ref{fig: sentence}.
\begin{figure}[h!]
      \centering
      \includegraphics[width=1\textwidth]{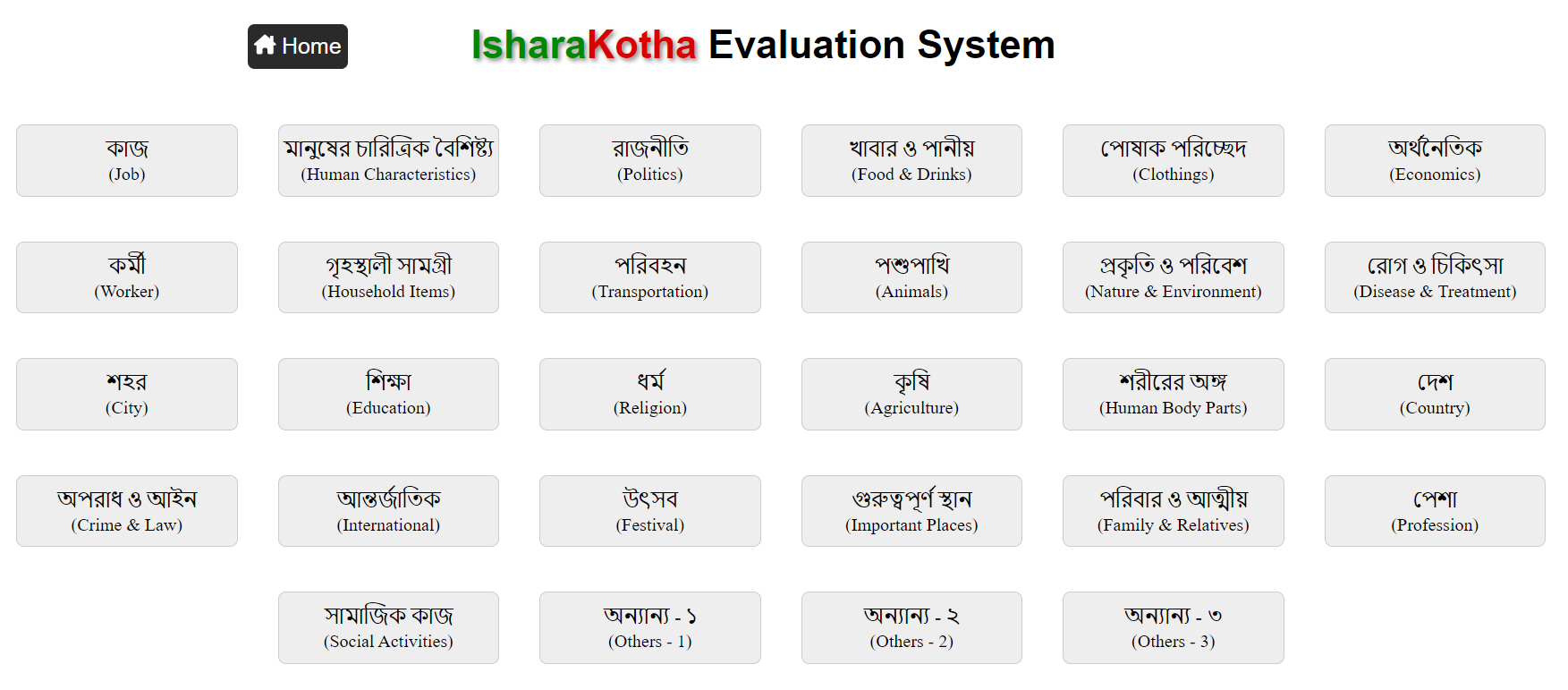}
      \caption{Word Categories in the Evaluation System}
      \label{fig: word categories}
\end{figure}

To visualize the animation using an avatar, an interpreter is required to select a specific word they wish to animate. The avatar then demonstrates the SL animation corresponding to the selected word. For feedback purposes, interpreters can assess the extent to which the animations effectively convey the meaning of the sign, using four predefined options: ``Bad", ``Average", ``Good" and ``Excellent". This rating system enables a quantifiable evaluation of the quality of the animations. Additionally, an optional comment box has been included to allow interpreters to offer specific suggestions or feedback on how the animation of a word can be enhanced. This feature encourages more detailed and constructive feedback for further improvement.
\begin{figure}[h!]
      \centering
      \includegraphics[width=0.95\textwidth]{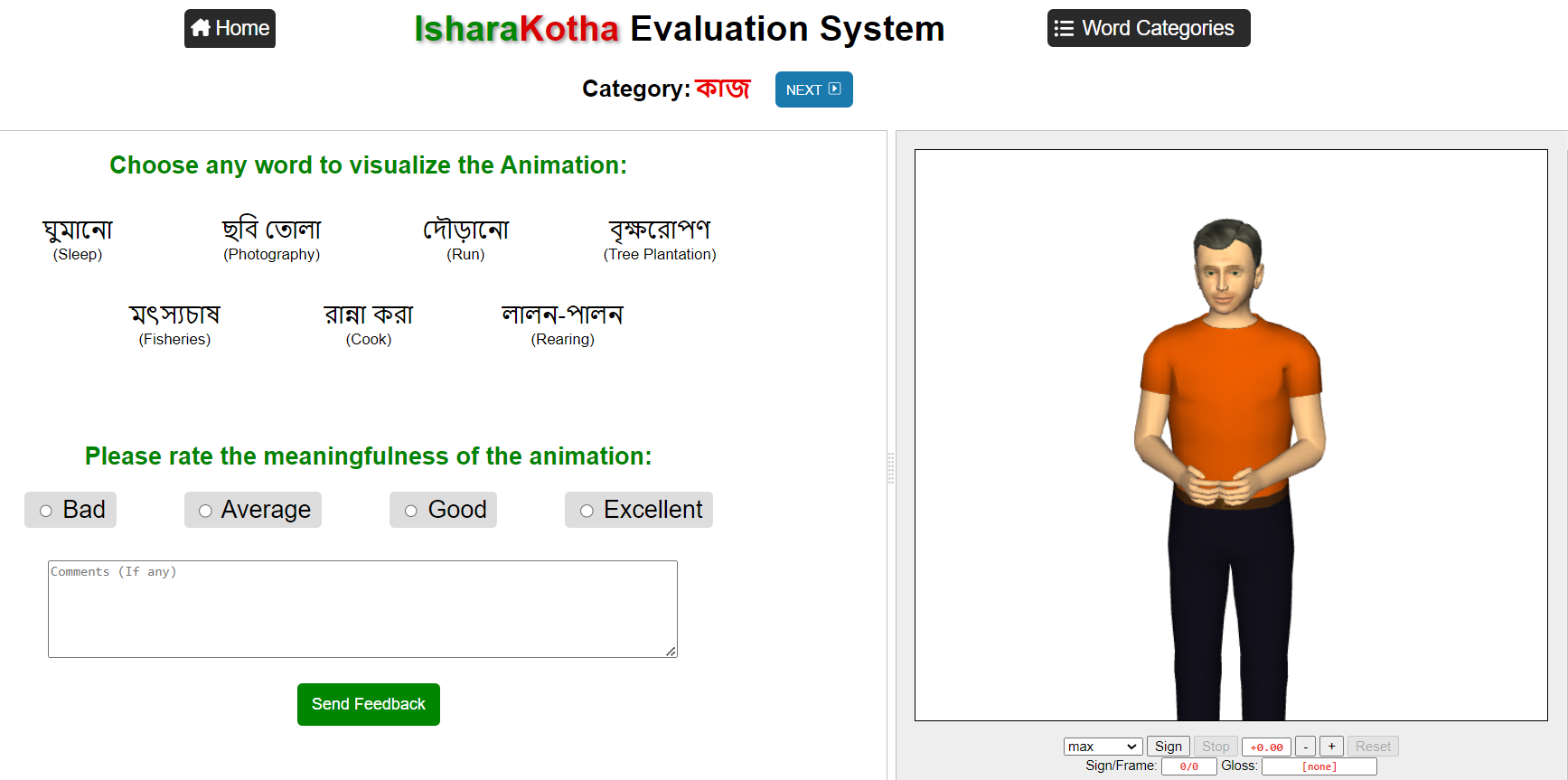}
      \caption{Evaluation of Sign Language Words}
      \label{fig: word}
\end{figure}
\begin{figure}[h!]
      \centering
      \includegraphics[width=0.97\textwidth]{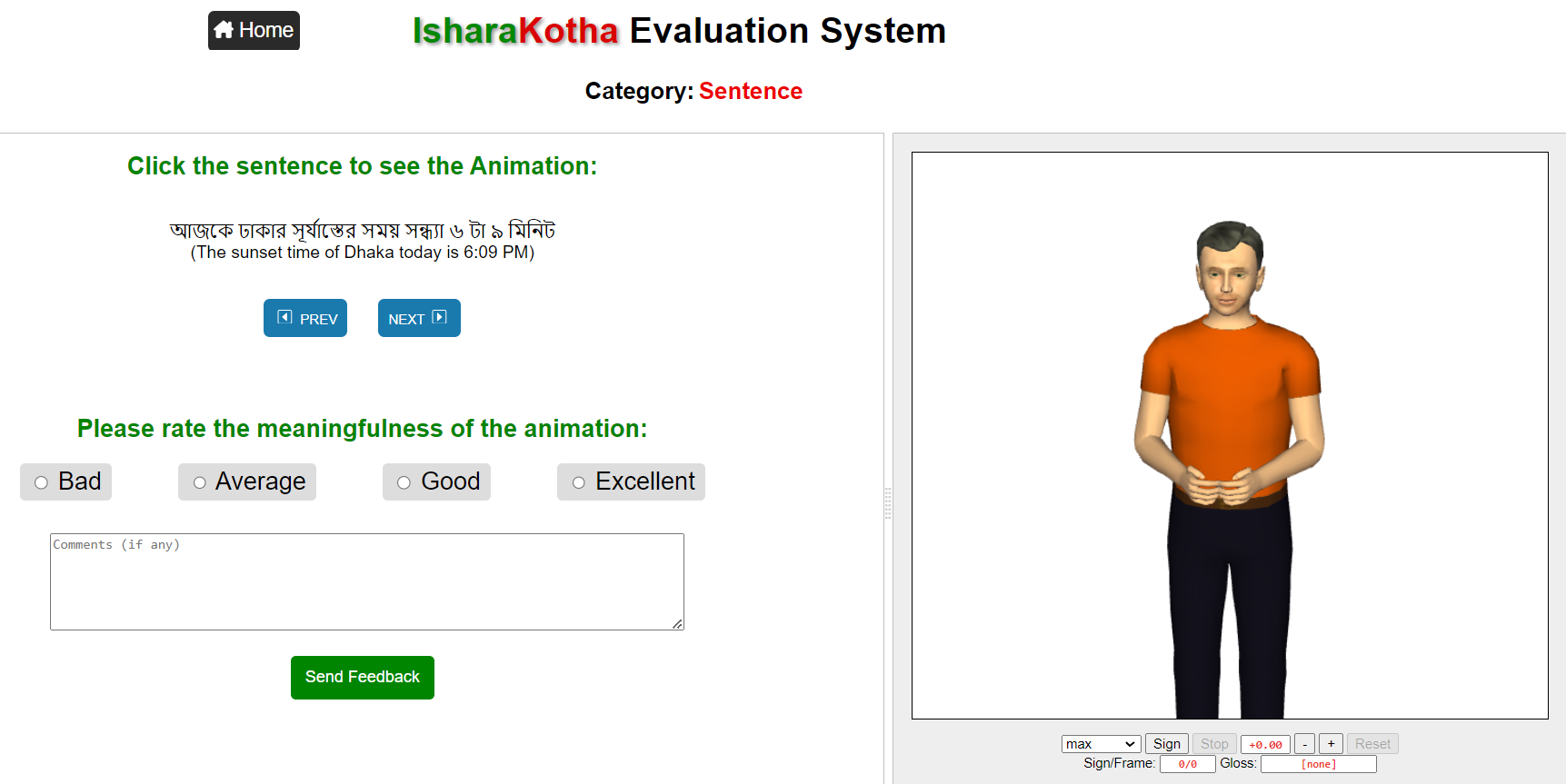}
      \caption{Evaluation of Sign Language Sentences}
      \label{fig: sentence}
\end{figure}

\section{Results and Discussion}
This section presents the evaluation results provided by two professional interpreters and one native sign language user. Our first interpreter \cite{tomaApu} assessed 30 alphabets, 10 numbers,  205 words, and 17 sentences, who provided ratings on the perceived meaningfulness of the sign animations. The interpreter was provided with four options for evaluation: ``Bad", ``Average", ``Good", and ``Excellent". Each option has been assigned with a specific numeric score. A score of 1, 2, 3, and 4 have been assigned for ``Bad", ``Average", ``Good" and ``Excellent" respectively. The results are summarized in Table \ref{Evaluation Score of 1st interpreter}. Let 
B represent the number of ``Bad" ratings, A the number of ``Average" ratings, G the number of ``Good" ratings, E the number of ``Excellent" ratings, and N the total number of ratings. The simplified formula for calculating the score is given by:
\[
\text{Score} = \frac{{B \times 1 + A \times 2 + G \times 3 + E \times 4}}{{N}}
\]
\newline
In the Alphabets section, out of 30 evaluations, 1 received a ``Bad" rating, 3 were considered ``Average", 26 were rated as ``Good" and none were rated ``Excellent." This indicates a generally positive perception of the clarity and accuracy of the alphabet animations. The average score for this section was calculated to be 2.83.

Moving on to the Numbers section, there were 10 evaluations in total. No animations were rated as ``Bad", 1 was rated ``Average", 2 received a ``Good" rating and 7 were considered ``Excellent". This suggests a high level of satisfaction with the representation of numerical symbols using avatar animations. The average score for this section was calculated to be 3.60.

In the Words section, a comprehensive evaluation was conducted across 205 instances randomly. Interpreters rated 4 animations as ``Bad", 16 as ``Average", 78 as ``Good" and 107 as ``Excellent". This indicates a generally positive perception of the quality and effectiveness of the word animations. The average score for this section was calculated to be 3.40.

In the Sentences section, a total of 17 evaluations were conducted. Among these evaluations, there was one rating of ``Bad", two ratings of ``Average", nine ratings of ``Good" and five ratings of ``Excellent". These ratings resulted in an average score of 3.06 for this section.

The \textit{IsharaKotha} corpus received a total of 6 ``Bad" ratings, 22 ``Average" ratings, 115 ``Good" ratings, and 119 ``Excellent" ratings, resulting in an average overall score of 3.32 out of 4 from the first interpreter.
\begin{table}[h!]
\centering
\caption{Evaluation Scores given by the first Interpreter}
\renewcommand{\arraystretch}{1.2}
\begin{tabular}{|c|c|c|c|c|c|} 
\hline
\textbf{Section Name} & \textbf{Bad} & \textbf{Average} & \textbf{Good} & \textbf{Excellent} & \textbf{Score}\\
\hline
Alphabets & 1 & 3 & 26 & 0 & 2.83\\
Numbers & 0 & 1 & 2 & 7 & 3.60\\
Words & 4 & 16 & 78 & 107 & 3.40\\
Sentences & 1 & 2 & 9 & 5 & 3.06\\
\hline
\centering \textbf{Total} & 6 & 22 & 115 & 119 & 3.32 \\
\hline
\end{tabular}
\label{Evaluation Score of 1st interpreter}
\end{table}

Similarly, reviewing the Alphabets section, the second interpreter provided 4 ``Bad'', 1 ``Average'', 2 ``Good'', and 5 ``Excellent'' ratings. In the Numbers section, 4 animations were rated as ``Good'' and 6 as ``Excellent'', with none received a rating of ``Bad'' or ``Average''. Additionally, 10 animations were rated as ``Bad'', 62 as ``Average'', 122 as ``Good'', and 108 as ``Excellent'' in the Words section. Lastly for the Sentences section, no animation was rated as ``Bad'', but 3 as ``Average'', 7 as ``Good'', and 9 were rated as ``Excellent''. Therefore, 3.17 out of 4 is the average assessment score based on second interpreter's feedback depicted in Table \ref{Evaluation Score of 2nd interpreter}.
\begin{table}[h!]
\centering
\caption{Evaluation Scores given by the second Interpreter}
\renewcommand{\arraystretch}{1.2}
\begin{tabular}{|c|c|c|c|c|c|} 
\hline
\textbf{Section Name} & \textbf{Bad} & \textbf{Average} & \textbf{Good} & \textbf{Excellent} & \textbf{Score}\\
\hline
Alphabets & 4 & 1 & 2 & 5 & 2.66\\
Numbers & 0 & 0 & 4 & 6 & 3.60\\
Words & 10 & 62 & 122 & 108 & 3.09\\
Sentences & 0 & 3 & 7 & 9 & 3.32\\
\hline
\centering \textbf{Total} & 14 & 66 & 135 & 128 & 3.17 \\
\hline
\end{tabular}
\label{Evaluation Score of 2nd interpreter}
\end{table}

In addition to two interpreters, one sign language user also evaluated our system. As demonstrated in Table \ref{Evaluation Score of 3rd interpreter}, the feedback indicated that the Alphabet section performed above average with 11 ``Bad'', 3 ``Average'', 26 ``Good'' and 3 as ``Excellent'' ratings. This section received an average score of 2.48 out of 4. With a perfect score of 4 out of 4, the Number section demonstrated outstanding performance and received overwhelmingly all favorable reviews. Words section achieved a great performance with an average score of 3.14 and received a good number of encouraging feedback. Finally, the Sentences section received an average score of 3.05, which denotes the majority of feedback being positive but with a few minor comments that might have been improved. However, there were given 3223 feedbacks in total, of which 96 were being rated as ``Bad'', 206 as ``Average'', 2096 as ``Good'', and 825 as ``Excellent'', resulting in an overall evaluation score of 3.13 out of 4 from the sign language user. 
\begin{table}[h!]
\newcolumntype{P}[1]{>{\centering\arraybackslash}p{#1}}
\centering
\caption{Evaluation Scores given by a sign language user}
\renewcommand{\arraystretch}{1.2}
\begin{tabular}{|c|c|c|c|c|c|} 
\hline
\textbf{Section Name} & \textbf{Bad} & \textbf{Average} & \textbf{Good} & \textbf{Excellent} & \textbf{Score}\\
\hline
Alphabets & 11 & 3 & 26 & 3 & 2.48\\
Numbers & 0 & 0 & 0 & 10 & 4.00\\
Words & 84 & 202 & 2056 & 808 & 3.14\\
Sentences & 1 & 1 & 14 & 4 & 3.05\\
\hline
\centering \textbf{Total} & 96 & 206 & 2096 & 825 & 3.13 \\
\hline
\end{tabular}
\label{Evaluation Score of 3rd interpreter}
\end{table}

These evaluation results demonstrate that the \textit{IsharaKotha} corpus has generally been well-received by interpreters and user, resulting the final evaluation score of 3.14 out of 4, as highlighted in Table \ref{Final Evaluation Score}. The high number of ``Good" and ``Excellent" ratings across all sections indicates a positive impression of the animation quality and effectiveness. However, there is scope for improvement in addressing the instances where the ratings were ``Bad" or ``Average", especially in the Alphabets section.
\begin{table}[h!]
\centering
\caption{Evaluation Score Summary}
\renewcommand{\arraystretch}{1.2}
\begin{tabular}{|c|c|c|c|c|c|} 
\hline
\textbf{} & \textbf{Bad} & \textbf{Average} & \textbf{Good} & \textbf{Excellent} & \textbf{Score}\\
\hline
First interpreter & 6 & 22 & 115 & 119 & 3.32\\
Second interpreter & 14 & 66 & 135 & 128 & 3.17\\
Sign language user & 96 & 206 & 2096 & 825 & 3.13\\
\hline
\centering \textbf{Total} & 116 & 294 & 2346 & 1072 & 3.14 \\
\hline
\end{tabular}
\label{Final Evaluation Score}
\end{table}

As shown in Table \ref{Comparison}, when compared to three notable existing datasets in terms of scope, methodology, and content, our comprehensive Bangla Sign Language dataset, \textit{IsharaKotha}, significantly expands the scope of BDSL resources. By incorporating a large number of word sign animations, it establishes itself as a valuable resource for researchers and practitioners.
\begin{table}[h!]
\centering
\caption{Comparison of Bangla Sign Language Datasets}
\renewcommand{\arraystretch}{1.3}
\begin{tabular}{|c|c|c|>{\centering\arraybackslash}m{3.2cm}|}
\hline
\textbf{Dataset} & \textbf{Content} & \textbf{Methodology} & \textbf{\centering Scope}\\
\hline
Islam et al. (2022) & 13 numeral gestures & HamNoSys and SiGML & Limited to numerals \\
\hline
Sams et al. (2023) & 6,000 sign videos & Video-based & 200 unique words \\
\hline
Hasib et al. (2022) & 29,490 images & Static images & 49 alphabets \\
\hline
IsharaKotha & 3,823 word sign animations & HamNoSys and SiGML & Alphabets, numerals and words\\
\hline

\end{tabular}
\label{Comparison}
\end{table}
\newpage
\section{Limitations}

Resource development and benchmarking are crucial concerns when dealing with a low-resource and morphologically complex language such as Bangla. The domain of Bangla sign language is equally complicated and diverse. While creating the \textit{IsharaKotha} corpus, we encountered difficulties in representing certain directional signs for some words. For instance, these difficulties included touching the lower parts of the body, touching the backside of body parts, pointing along the spine, and representing some real-time scenario-based facial expressions. These specific signs may require special handling. Moreover, our current system assumes that all root words are present in the sign dictionary. Generating dynamic animations for any Bangla sentence requires mapping the entire vocabulary to the available 3823 signs, a process that is not yet complete.
\begin{figure}[h]
    \centering
    \subfloat{
        \includegraphics[scale=0.31]{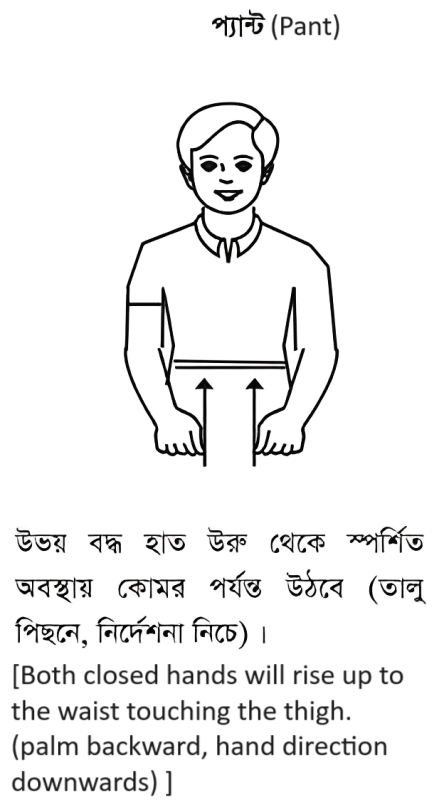}
    }
    \subfloat{
        \includegraphics[scale=0.24]{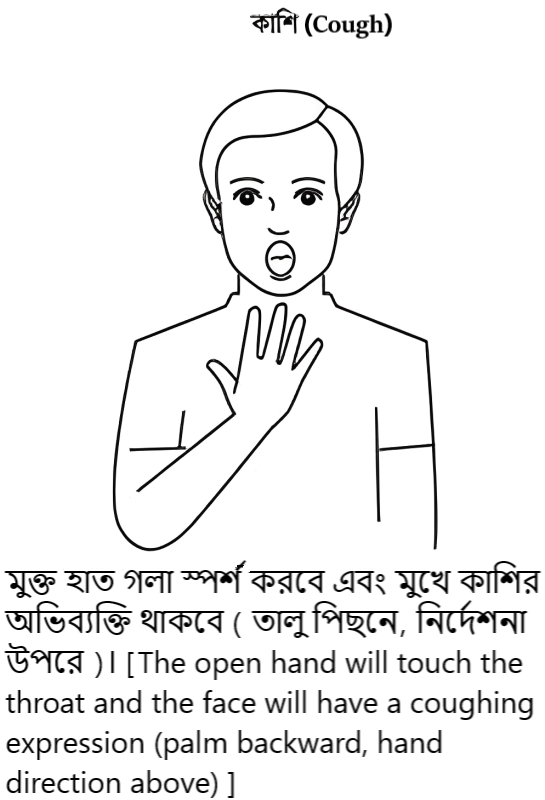}
    }\hspace{0.05cm}
    \subfloat{
        \includegraphics[scale=0.24]{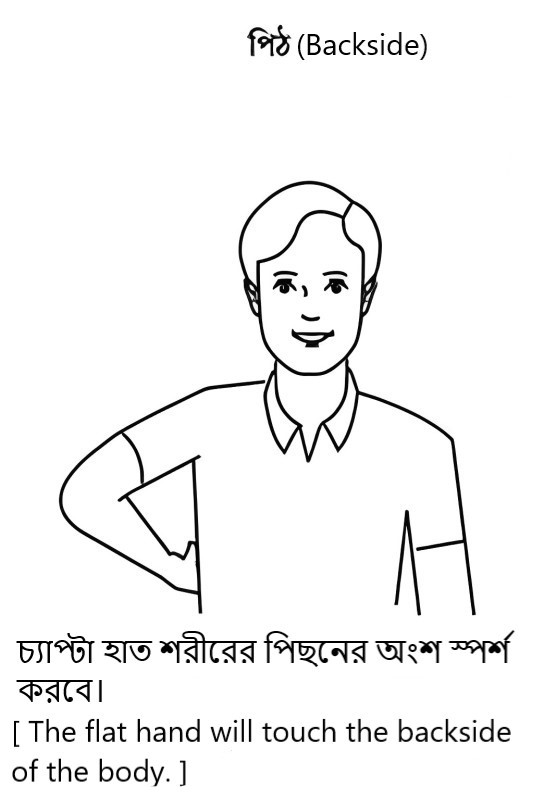}
    }

    \caption{Examples of some unfinished signs from the BDSL Dictionary}
    \label{Dataset}
\end{figure}

\section{Conclusion}
Sign Language generation systems have been successfully developed in many other countries. However, due to the absence of an appropriate corpus, similar systems have yet to be developed for Bangla language. In this article, we introduced \textit{IsharaKotha}, the first HamNoSys-based Bangla Sign Language corpus. Unlike pre-recorded video-based systems, this approach is memory-efficient, eliminates the need to hire volunteers, and is less time-consuming. A sign language corpus is vital for raising awareness and fostering understanding of sign language among the general public. This resource is expected to play a pivotal role in education, research, and enhancing communication accessibility. A well-developed corpus not only facilitates the learning and study of Sign Language but also supports the creation of tools and technologies that bridge the communication divide between the hearing-impaired and the broader community. This ensures that everyone has an equal opportunity to engage in different aspects of society. In the future, we aim to develop a dynamic translation system that converts Bangla text into sign language animations. This avatar-based sign language translation system can also be integrated with other communication technologies, such as speech recognition, to enable more comprehensive and natural interactions.

\section{Acknowledgment}
The authors express their gratitude to the Research Center of Shahjalal University of Science and Technology
for providing partial funding for this research.


\bibliographystyle{elsarticle-num} 
\bibliography{bibFileReference}





\end{document}